\documentclass{ws-ijmpa}

\begin{document}
\def\gappeq{\mathrel{\rlap {\raise.5ex\hbox{$>$}} {\lower.5ex\hbox{$\sim$}}}}
\def\lappeq{\mathrel{\rlap{\raise.5ex\hbox{$<$}} {\lower.5ex\hbox{$\sim$}}}}

\markboth{GUIDO ALTARELLI}{Concluding Talk: QCD 2005}

\catchline{}{}{}{}{}

\title{Concluding Talk: QCD 2005}

\author{GUIDO ALTARELLI}
\address{ CERN, Department of Physics, Theory Division \\  
CH-1211 Gen\`eve 23, Switzerland\\
and\\
Dipartimento di Fisica, Universita' di Roma Tre\\
Rome, Italy\\
 {\rm E-mail: guido.altarelli@cern.ch}}


\maketitle

\begin{abstract}
This is neither a summary talk (too much for too short a talk) nor a conclusion (a gigantic work is in progress and we are not at the end of a particular phase), rather an overview of the field as reflected at this Conference.
\end{abstract}

\keywords{particle physics; strong interactions; chromodynamics}

\vskip 1.5cm

QCD stands today as a main building block of the Standard Model (SM) of particle physics. For many years the relativistic quantum field theory of reference was QED, but now QCD offers a much more complex and intriguing theoretical laboratory (an historical perspective was offered by Applequist in his fine introduction). Due to asymptotic freedom, in fact QCD is a better defined theory than QED. The statement that QCD is an unbroken renormalisable gauge theory with six kinds of triplets quarks with given masses completely specifies the form of the Lagrangian in terms of quark and gluon fields. From the compact form of its Lagrangian one might be led to  think that QCD is a "simple" theory. But actually this simple theory has an extremely rich dynamical content, including the striking property of confinement,  a complex hadron spectrum (with light and heavy quarks), the spontaneous breaking of (approximate) chiral symmetry, a complicated phase transition structure (deconfinement, chiral symmetry restauration, colour superconductivity), a highly non trivial vacuum topology (instantons, $U(1)_A$ symmetry breaking, strong CP violation,....), the property of asymptotic freedom and so on.

How do we get predictions from QCD? There are non perturbative methods: lattice simulations (in great continuous progress), effective lagrangians valid in restricted specified domains [chiral lagrangians, heavy quark effective theories, Soft Collinear Effective Theories (SCET), Non Relativistic QCD....reviewed at this Conference in the talks by Maussallam, Bauer, Bodwin...] and also QCD sum rules, potential models (for quarkonium) and so on. The perturbative approach, based on asymptotic freedom and valid for hard processes, still remains the main quantitative connection to experiment.

Due to confinement no free coloured particles are observed but only colour singlet hadrons. In high energy collisions the produced quarks and gluons materialize as narrow jets of hadrons. The understanding of the confinement mechanism has much improved over the years thanks to lattice simulations of QCD at finite temperatures and densities. The progress in this area has been reviewed in the talks by Satz, Hatsuda, Aoki and Kuti. The potential between two colour charges clearly shows a linear slope at large distances (linearly rising potential). The slope decreases with increasing temperature until it vanishes at a critical temperature $T_C$. Above $T_C$ the slope remains zero. The phase transitions of colour deconfinement and of chiral restauration appear to happen together on the lattice. A rapid transition is observed in lattice simulations where the energy density $\epsilon(T)$ is seen to sharply increase near the critical temperature for deconfinement and chiral restauration. The critical parameters and the nature of the phase transition depend on the number of quark flavours $N_f$ and on their masses. For example, for  $N_f$ = 2 or 2+1 (i.e. 2 light u and d quarks and 1 heavier s quark), $T_C \sim 175~MeV$  and $\epsilon(T_C) \sim 0.5-1.0 ~GeV/fm^3$. For realistic values of the masses $m_s$ and $m_{u,d}$ the phase transition appears to be a second order one, while it becomes first order for very small or very large $m_{u,d,s}$. The hadronic phase and the deconfined phase are separated by a crossover line at small densities and by a critical line at high densities. Determining the exact location of the critical point in T and $\mu_B$ is an important challenge for theory which is also important for the interpretation of heavy ion collision experiments. At high densities the colour superconducting phase is also present with bosonic diquarks acting as Cooper pairs. 

A large investment is being done in experiments of heavy ion collisions with the aim of finding some evidence of the quark gluon plasma phase. Many exciting results have been found at the CERN SPS in the past  years and more recently at RHIC. At the CERN SPS some experimental hints of variation with the energy density were found in the form, for example, of $J/ \Psi$ production suppression or of strangeness enhancement when going from p-A to Pb-Pb collisions. Indeed a posteriori the CERN SPS appears well positioned in energy to probe the transition region, in that a marked variation was observed for different observables. The results from RHIC and the status of the experimental search for the quark-gluon plasma have been reviewed here in the talks by McLerran, Qiu and Wang. The most impressive effect detected at RHIC, interpreted as due to the formation of a hot and dense bubble of nuclear matter, is the observation of a strong suppression of back-to-back correlations in jets from central collisions in Au-Au, showing that the jet that crosses the bulk of the dense region is absorbed. However, it is fair to say that the significance of each single piece of evidence for a change of regime can be questioned and one is still far from an experimental confirmation of a phase transition. The experimental programme on heavy ion collisions will continue at RHIC and then at the LHC where ALICE, a dedicated heavy ion collision experiment, is in preparation.

As we have seen the main approach to non perturbative problems in QCD is by simulations of the theory on the lattice, a technique started by K. Wilson in 1974 which has shown continuous progress over the last decades. The status of lattice QCD has been discussed at this Conference in the talks by Kronfeld, Negele, Neuberger and Ukawa. One recent big step, made possible by the availability of more powerful dedicated computers, is the evolution from quenched (i.e. with no dynamical fermions) to unquenched calculations. In doing so an evident improvement in the agreement of predictions with the data is obtained. For example, modern unquenched simulations reproduce the hadron spectrum quite well. Calculations with dynamical fermions (which take into account the effects of virtual quark loops) imply the evaluation of the quark determinant which is a difficult task. How difficult depends on the particular calculation method. There several approaches (Wilson, twisted mass,  Kogut-Susskind staggered, Ginsparg-Wilson fermions), each with its own advantages and disadvantages. Another area of progress is the implementation of chiral extrapolations: lattice simulation is limited to large enough masses of light quarks. To extrapolate the results down to the physical pion mass one can take advantage of chiral 
effective theory results in order to control the chiral logs: $\log(m_q/4\pi f_\pi)$. Kronfeld emphasized that for lattice QCD one is now in an epoch of pre-dictivity as opposed to the  post-dictivity of the past. And in fact the range of precise lattice results currently includes many domains:  the QCD coupling constant (the value $\alpha_s=0.1177(13)$ was quoted by Kronfeld), the quark masses, the form factors for K and D decay, the B parameter for kaons, the decay constants $f_K$, $f_D$, $f_{Ds}$, the $B_c$ mass, the nucleon axial charge $g_A$ (as discussed by Negele, the lattice result is close to the experimental value $g_A \sim 1.25$ and well separated from the $SU(6)$ value $g_A = 5/3$) and many more.

QCD is playing a crucial role in the interpretation of experiments at B factories (see the talk by Bernard) by a combination of effective theory methods (heavy quark effective theory, NRQCD, SCET), lattice simulations and perturbative calculations, as described in the talks by Bauer, Beneke, Cheng and Du. Overall, B mixing and CP violation agree very well with the SM predictions based on the CKM matrix. As discussed by Fleischer it is only in channels that are forbidden at tree level and only go through penguin loops that some deviation is perhaps indicated (as is the case for $B \rightarrow  \pi K$ modes).

New developments in hadron spectroscopy have been discussed by Klempt, Jin and Nakano. Ordinary hadrons are baryons $B\sim qqq$ and mesons $M\sim q \bar q$. For a long time the search for exotic states was concentrated on glueballs, gg bound states, predicted  at $M\gappeq 1.5 ~GeV$ by the lattice. As well known, experimentally glueballs were never clearly identified, probably because they are largely mixed with states made up of quark-antiquark pairs. Hybrid states ($q \bar q g$ or $qqqg$) have also escaped detection. Recently a number of unexpected results have revamped the interest for hadron spectroscopy. Several experiments have reported new narrow states, with widths below a few MeV(!!): 
$\Theta^+(1540)$ with the quantum numbers of $nK^+$ or $pK^0_S$ or, in terms of quarks, of $uudd \bar s$; $D_{sJ}^+(2317) \sim D_s\pi$, $D_{sJ}^+(2460) \sim D^*_s\pi$,.... and $X^0(3872) \sim \pi \pi J/\Psi$. The interpretations proposed are in terms of pentaquarks ($[ud][ud]\bar s$ for $\Theta^+$ for example), tetraquarks ($[qq][\bar q \bar q]$) vs meson-meson molecules  for low lying scalar mesons and for 
$X^0$ and also in terms of chiral solitons. Tetraquarks and pentaquarks are based on diquarks: $[qq]$
of spin $0$, antisymmetric in colour, $\bar 3$ of $SU(3)_{colour}$, and antisymmetric in flavour, $\bar 3$ of $SU(3)_{flavour}$. Tetraquarks were originally proposed for scalar mesons by Jaffe. It is well known that there are two clusters of scalar mesons: one possible nonet at high mass, around $1.5~GeV$, and a low lying nonet below $1~GeV$. The light nonet presents an inversion in the spectrum: the mesons that would contain  s-quarks in the conventional $q\bar q$ picture and would hence be heavier are actually lighter. This becomes clear if the s quark with index "3" in the conventional picture is replaced be the diquark $[ud]$ as in the tetraquark interpretation. However, in his talk Pennington has raised doubts about the existence of so many scalar states. The tetraquark interpretation for the doubly charmed $X^0(3872)$ has been proposed recently by Maiani et al as opposed to that in terms of a $D-D*$ molecule by Braaten and Kusunoki. In his talk Bernard has pointed out that both models face difficulties in the data. For possible pentaquark states like the $\Theta^+$ doubts on their existence are relevant. Not only there are mass inconsistencies among different experiments,  evident tension between a small width and large production rates and the need of an exotic production mechanism to explain the lack of evidence at larger energies. But the most disturbing fact is the absence of the signal in some specific experiments where it is difficult to imagine a reason for not seeing it. The last, very troubling example, is the negative result, made public last April, from the CLAS g11 experiment at JLAB where no $\Theta^+$ was observed in a dedicated high statistics search. In conclusion the picture is not clear and more work is needed.

We consider now  perturbative QCD. In the QCD Lagrangian quark masses are the only parameters with dimensions. Naively (or classically) one would expect massless QCD to be scale invariant so that dimensionless observables would not depend on the absolute energy scale but only on ratios of energy variables. While massless QCD in the quantum version, after regularisation and renormalisation, is finally not scale invariant, the theory is asymptotically free and all the departures from scaling are asymptotically small, logarithmic and computable in terms of the running coupling $\alpha_s(Q^2)$. Mass corrections present in the realistic case as well as hadronisation effects are suppressed by powers. The QCD beta function that fixes the running coupling is known in QCD up to 4 loops in the $MS$ or $\bar{MS}$ definitions. The 4-loop calculation  by van Ritbergen, Vermaseren and Larin ('97) involving about 50.000 4-loop diagrams is a great piece of work. The running coupling is a function of $Q^2/\Lambda^2_{QCD}$, where $\Lambda_{QCD}$ is the scale that breaks scale invariance in massless QCD. Its value in $\bar{MS}$, for 5 flavours of quarks, from the PDG'04 is $\Lambda_{QCD}\sim 218(24)~MeV$. This fundamental constant of nature, which determines the masses of light hadrons, arises as a subtle effect from defining the quantum theory. There is no hierarchy problem in QCD, in that the logarithmic evolution of the running makes the smallness of $\Lambda_{QCD}$ with respect to the Planck mass $M_{Pl}$ natural: $\Lambda_{QCD}\sim M_{Pl} \exp{[-1/2b\alpha_s(M_{Pl}^2)]}$. 

The measurements of $\alpha_s(Q^2)$  are among the main quantitative tests of the theory. Since this important subject has not been covered at this meeting I will discuss it here with a minimum of details. The most  precise and  reliable determinations are from $e^+e^-$ colliders (mainly at LEP: inclusive hadronic Z decay, inclusive
hadronic $\tau$ decay, event shapes and jet rates) and from Deep Inelastic Scattering (DIS) scaling violations.  

Z decay widths are very clean: the perturbative expansion is known to 3-loops, power  corrections are controlled by the light-cone operator expansion and are very suppressed due to $m_Z$ very large. The basic quantity is $\Gamma_h$ the Z hadronic partial width. It enters in $R_l$, $\sigma_h$, $\sigma_l$ and $\Gamma_Z$ (the width ratio of hadrons to leptons, the hadron cross section at the peak, the charged lepton cross section at the peak and the total width, respectively) which are separately measured with largely independent systematics. From combining all these measurements one obtains $\alpha_s(m^2_Z)= 0.1190(27)$ (see the LEPEWWG web page). The error is predominantly theoretical and is dominated by our ignorance on $m_H$ and from higher orders in the QCD expansion (the possible impact of new physics is very limited, given the results of precision tests of the SM at LEP).  The measurement of $\alpha_s(m_Z)$ from $\tau$ decay is based on $R_\tau$, the ratio of the hadronic to leptonic widths. $R_\tau$ has a number of advantages that, at least in part, tend to compensate for the smallness of $m_\tau$. First, $R_\tau$ is maximally inclusive, more than $R_{e^+e^-}(s)$, because one also integrates over all values of the invariant hadronic squared mass. Analyticity is used to transform the integral into one on the circle at $|s|=m_\tau^2$.  Also, a factor $(1-\frac{s}{m_\tau^2})^2$ that appears in the integral kills the sensitivity of the region $\rm{Re}s=m_\tau^2$ where the physical cut and the associated thresholds are located. Still the quoted result (PDG'04) looks a bit too precise: $\alpha_s(m_Z^2)=0.1210(7(exp))(30(th))$.
This precision is obtained by taking for granted that corrections suppressed by $1/m_\tau^2$ are negligible. This is
because, in the massless theory, in the light cone expansion there are no dim-2 Lorentz and gauge invariant operators. In
the massive theory, the coefficient of $1/m_\tau^2$ does not vanish but is proportional to light quark mass-squared $m^2$. This is still negligible if $m$ is taken as a Lagrangian mass of a few $MeV$. But would not at all be negligible, actually would much increase the theoretical error, if it is taken as a constituent mass of order $m\sim\Lambda_{QCD}$. Most people believe the optimistic version. I am not convinced that the gap is not filled up by ambiguities of $0(\Lambda_{QCD}^2/m_\tau^2)$ e.g. from ultraviolet renormalons. In any case, one can discuss the error, but it is true and remarkable, that the central value from $\tau$ decay, obtained at very small $Q^2$, when evolved at $Q^2=m_Z^2$, is in perfect agreement with all other precise determinations of $\alpha_s(m_Z^2)$ at more typical LEP values of $Q^2$, offering a direct evidence for the running. The measurements of $\alpha_s$ from event shapes and jet rates are affected by non perturbative hadronic corrections which are difficult to precisely assess. The combined result gives $\alpha_s(m_Z^2)=0.120(6)$ (PDG'04).  By measuring event shapes at different energies in the LEP1 and LEP2 ranges one also directly sees the running of $\alpha_s$.

In DIS QCD predicts the $Q^2$ dependence of a generic structure function $F(x,Q^2)$ at each fixed x, not the x shape. But the $Q^2$ dependence is related to the x shape by the QCD evolution equations. For each x-bin the data allow to extract the slope of an approximately stright line in $dlogF(x,Q^2)/dlogQ^2$: the log slope. The $Q^2$ span and the precision of the data are not much sensitive to the curvature, for most x values. A single value of $\Lambda_{QCD}$ must be fitted to reproduce the collection of the log slopes. The QCD theory of scaling violations, based on the renormalization group and the light-cone operator expansion, is crystal clear. Recently ('04) the formidable task of computing the splitting functions at NNLO accuracy has been completed by Moch, Vermaseren and Vogt, a really monumental, fully analytic calculation. For the determination of $\alpha_s$ the scaling violations of non-singlet structure functions would be ideal, because of the minimal impact of the choice of input parton densities. Unfortunately the data on non-singlet structure functions are not very accurate. For example, NNLO determinations of $\alpha_s$ from the CCFR data on $F_{3\nu N}$ with different techniques have led to the central values $\alpha_s(m_Z^2)=0.1153$ (Santiago and Yndurain '01), $\alpha_s(m_Z^2)=0.1174$ (Maxwell, Mirjalili '02), $\alpha_s(m_Z^2)=0.1190$ (Kataev et al '02), with average and common estimated error of $\alpha_s(m_Z^2)=0.117(6)$ which I will use later. When one measures $\alpha_s$ from scaling violations on $F_2$ from e or $\mu$ beams, the data are abundant, the errors small but there is an increased dependence on input parton densities and especially a strong correlation between the result on $\alpha_s$ and the input on the gluon density. There are several most complete and accurate derivations of $\alpha_s$ from scaling violations in $F_2$ with different, sophisticated methods (Mellin moments, Bernstein moments, truncated moments....).  We quote here the result at NNLO accuracy from MRST'03 (see PDG'04): $\alpha_s(m_Z^2)=0.1153(36)$.

More measurements of $\alpha_s$ could be listed: I just reproduced those which I think are more significant and reliable. There is a remarkable agreement among the different determinations. If I directly average the five values from inclusive Z decay, from $R_\tau$, from event shapes and jet rates in $e^+e^-$, from $F_3$ and from $F_2$ in DIS I obtain $\alpha_s(m_Z^2)=0.1188(16)$ in perfect match with the PDG'04 average $\alpha_s(m_Z^2)=0.1187(20)$.

The importance of DIS for QCD goes well beyond the measurement of $\alpha_s$. In the past it played a crucial role in establishing the reality of quarks and gluons as partons and in promoting  QCD as the theory of strong interactions. Nowadays it still generates challenges to QCD as, for example, in the domain of structure functions at small x (see the talks by Bartels and by Forte) or of polarized structure functions (Saito) or of generalized parton densities (Schaefer) and so on. 

The problem of constructing a convergent procedure to include the BFKL corrections at small x in the singlet splitting functions, in agreement with the small-x behaviour observed at HERA, has been a long standing puzzle which has now been essentially solved, as described by Forte. The naive BFKL rise of splitting functions is tamed by resummation of collinear singularities and by running coupling effects. The resummed expansion is well behaved and the result is close to the NLO perturbative splitting function in the region of HERA data at small x. 

In polarized DIS one main question is how the proton helicity is distributed among quarks, gluons and orbital angular momentum: $1/2\Delta \Sigma + \Delta g + L_z= 1/2$.
The quark moment $\Delta \Sigma$ was found to be small: typically $\Delta \Sigma_{exp} \sim 0.2$ (the "spin crisis"). Either $\Delta g + L_z$ is large or there are contributions to $\Delta \Sigma$ at very small x outside of the measured region. $\Delta g$ evolves like $\Delta g \sim log Q^2$, so that eventually should become large (while $\Delta \Sigma$ and $\Delta g + L_z$ are $Q^2$ independent in LO). It will take long before this log growth of $\Delta g$ will be confirmed by experiment! $\Delta g$ can be measured indirectly by scaling violations and directly from asymmetries, e.g. in $c \bar c$ production. Existing direct measurements by Hermes, Compass, and at RHIC are still very crude and show no hint of a large $\Delta g$. The perspectives of better measurements are good at Compass and RHIC in the near future.

Another important role of DIS is to provide information (see the talk by Tung) on parton density functions (PDF) which are instrumental for computing cross-sections of hard processes at hadron colliders via the factorisation formula. The predictions for cross sections and distributions at $pp$ or $p\bar p$ colliders for large $p_T$ jets or photons, for heavy quark production, for Drell-Yan, W and Z production are all in very good agreement with experiment. There was an apparent problem for b quark production at the Tevatron, but the problem appears now to be solved by a combination of refinements (log resummation, B hadrons instead of b quarks, better fragmentation functions....). The QCD predictions are so solid that W and Z production are actually considered as possible luminosity monitors for the LHC. 

A great effort is being devoted to the preparation for the LHC. Calculations for specific processes are being completed. A very important example is Higgs production via $g ~+~ g \rightarrow H$ (see the talk by Li). The amplitude is dominated by the top quark loop. Higher order corrections can be computed either in the effective lagrangian approach, where the heavy top is integrated away and the loop is shrunk down to a point [the coefficient of the effective vertex is known to $\alpha_s^4$ accuracy (Chetyrkin at al)], or in the full theory. At the NLO the two approaches agree very well for the rate as a function of $m_H$. Rapidity and $p_T$ distributions have also been evaluated at NLO. The $[log(p_T/m_H)]^n$ have been resummed in analogy with what was done long ago for W and Z production. Recently the NNLO analytic calculation for the rate has been completed in the effective lagrangian formalism (Harlander and Kilgore, Ravindran et al, Anastasiou and Melnikov). 

The activity on event simulation also received a big boost from the LHC preparation. General algorithms for performing NLO calculations numerically  (requiring techniques for the cancellation of singularities between real and virtual diagrams), for example the dipole formalism by Catani, Seymour et al. The matching of matrix element calculation of rates together with the modeling of parton showers has been realised in packages, as for example in the MC@NLO based on HERWIG (see the talk by Soper). The matrix element calculation, improved by resummation of large logs, provides the hard skeleton (with partons at large $p_T$) while the parton shower is constructed for each parent parton by a sequence of factorized collinear emissions fixed by the QCD splitting functions. In addition, at low scales a model of hadronisation completes the simulation. The importance of all the components, matrix element, parton shower and hadronisation can be appreciated in simulations of hard events compared with the Tevatron data. 

Before closing I would like to mention some very interesting developments at the interface between string theory and QCD, twistor calculus. A precursor work was the Parke-Taylor result in '86 on the amplitudes for n incoming gluons with given ± helicities. Inspired by dual models, they derived a compact formula for the maximum non vanishing helicity violating amplitude (with n-2  plus and 2 minus helicities) in terms of spinor products. Using the relation between strings and gauge theories in twistor space Witten developed in '03 a formalism in terms of effective vertices and propagators that allows to compute all helicity amplitudes. The method, much faster than Feynman diagrams, leads to very compact results. Since then rapid progress followed: for tree level processes powerful recurrence relations were established (Britto, Cachazo, Feng; Witten), the method was extended  to include massless fermions (Georgiou, V. Khoze) and also external EW vector bosons (Bern et al)  and Higgs particles (Dixon, Glover, Khoze, Badger et al). The level already attained is already important for multijet events at the LHC. And the study of loop diagrams has been started. In summary, this road looks very promising.

In conclusion, I think that the field of QCD as it was reviewed at this Conference appears as one of great maturity but also of robust vitality with many rich branches and plenty of new blossoms. The physics content of QCD is very large and our knowledge, especially in the non perturbative domain, is still very limited but progress both from experiment (LEP, HERA, Tevatron, RHIC, LHC......) and from theory is continuing at a healthy rate. And all the QCD predictions that we were able to formulate and to test are in very good agreement with experiment.


\section*{Acknowledgements}

As a last speaker it is a pleasant duty for me, on behalf of all the participants, to most warmly congratulate and thank the Local Organisers for this perfectly run meeting in this tantalizing city. We all enjoyed the atmosphere of exploding development that one immediately perceives here in Beijing which impressed so much all of us. And this is  also true for those, like me, who are not new to China, but yet found a surprisingly different environment. In particular my personal thankful compliments go to Chuan Liu and Xiangdong Ji and to all the Peking University Staff and the graduate students who have been so kind to me.



\end{document}